\documentclass{llncs}
\usepackage[T1]{fontenc}
\usepackage[utf8x]{inputenx}
\usepackage[british]{babel}
\usepackage{graphicx}
\usepackage{xspace}
\usepackage{amsmath}
\usepackage{amssymb}
\usepackage{multirow}
\usepackage{url}
\usepackage{hyperref}
\usepackage{cleveref}
\usepackage{tikz}
\usepackage{enumitem}
\usepackage{proof}

\usepackage{pifont}
\newcommand{\cmark}{\ding{51}}
\newcommand{\xmark}{\ding{55}}

\setitemize{noitemsep,topsep=0pt,parsep=0pt,partopsep=0pt,leftmargin=11.5pt}
\crefname{figure}{fig.}{figs.}
\Crefname{figure}{Fig.}{Figs.}
\crefname{section}{sect.}{sects.}
\Crefname{section}{Sect.}{Sects.}
\crefname{algocf}{alg.}{algs.}
\Crefname{algocf}{Alg.}{Algs.}
\crefname{equation}{eq.}{eqs.}
\Crefname{equation}{Eq.}{Eqs.}

\usepackage{tikz}
\usetikzlibrary{calc,automata,positioning,decorations.pathreplacing}
\tikzset{align at top/.style={baseline=(current bounding box.north)}}
\tikzstyle{every node}=[font=\scriptsize]
\tikzstyle{state} = [draw,fill=white,circle,thick,align=center,inner sep=0pt,minimum size=4.5mm]
\tikzstyle{lstate} = [draw,fill=white,rectangle,rounded corners,thick,align=center,inner sep=2pt]
\tikzstyle{dot} = [fill,circle,inner sep=0mm,minimum size=1.25mm,line width=0mm]

\makeatletter
\DeclareRobustCommand{\rvdots}{%
  \vbox{
    \baselineskip4\p@\lineskiplimit\z@
    \kern-\p@
    \hbox{.}\hbox{.}\hbox{.}
  }}
\makeatother

\usepackage[vlined,linesnumbered]{algorithm2e}
\SetAlCapFnt{\small}
\SetAlCapSkip{2ex plus 0.5ex minus 1ex}
\SetAlgoCaptionSeparator{.}
\SetEndCharOfAlgoLine{}
\SetArgSty{textrm}
\SetCommentSty{textit}
\SetKw{Break}{break}
\SetKwProg{With}{with}{:}{end}
\SetKwProg{fn}{function}{}{}
\SetKwFor{WhileTrue}{repeat}{}{end}
\SetKw{Continue}{continue}
\SetKwRepeat{DoWhile}{repeat}{while}
\crefname{algocf}{algorithm}{algorithms}
\Crefname{algocf}{Algorithm}{Algorithms}
\SetKwFor{ForEach}{foreach}{do}{}
\SetKwProg{Fn}{function}{}{}

\newcommand{\eg}{e.g.\ }
\newcommand{\ie}{i.e.\ }
\newcommand{\st}{\ifmmode \:\text{s.t.}\ \else s.t.\xspace\fi}

\newcommand{\wrt}{w.r.t.\xspace}
\newcommand{\sunit}[1]{\text{\begin{scriptsize}\,#1\end{scriptsize}}}
\newcommand{\funit}[2]{\text{\begin{scriptsize}\,\ensuremath{\frac{\text{#1}}{\text{#2}}}\end{scriptsize}}}
\newcommand{\ssunit}[1]{\text{\begin{scriptsize}#1\end{scriptsize}}}
\newcommand{\prism}{\mbox{\textsc{Prism}}\xspace}

\newcommand{\toolset}{\mbox{\textsc{Modest}} \textsc{Toolset}\xspace}
\newcommand{\mcsta}{\mbox{\textsc{mcsta}}\xspace}
\newcommand{\alg}[1]{\textsf{#1}\xspace}

\renewcommand{\implies}{\ensuremath{\mathrel{\Rightarrow}}\xspace}

\newcommand{\Dist}[1]{\ensuremath{\mathrm{Dist}({#1})}\xspace}
\newcommand{\support}[1]{\ensuremath{\mathrm{support}({#1})}\xspace}
\newcommand{\set}[1]{\ensuremath{\{\,#1\,\}}}
\newcommand{\Dirac}[1]{\ensuremath{\mathcal{D}(#1)}\xspace}
\newcommand{\tuple}[1]{\ensuremath{\langle #1 \rangle}}
\newcommand{\powerset}[1]{\ensuremath{2^{#1}}\xspace}
\newcommand{\NN}{\ensuremath{\mathbb{N}}\xspace}

\newcommand{\RR}{\ensuremath{\mathbb{R}}\xspace}

\newcommand{\QQplus}{\ensuremath{\mathbb{Q}^+}\xspace}

\newcommand{\RRpluszero}{\ensuremath{\mathbb{R}^{+}_{0}}\xspace}

\newcommand{\States}{\ensuremath{S}\xspace}
\newcommand{\InitialState}{\ensuremath{s_\mathit{init}}\xspace}
\newcommand{\Alphabet}{\ensuremath{A}\xspace}
\newcommand{\Trans}{\ensuremath{T}\xspace}
\newcommand{\defeq}{\mathrel{\vbox{\offinterlineskip\ialign{\hfil##\hfil\cr{\tiny \rm def}\cr\noalign{\kern0.30ex}$=$\cr}}}}

\newcommand{\xtr}[1]{\xrightarrow{\protect{\raisebox{-1.5pt}[0pt][0pt]{\ensuremath{\scriptstyle{#1}}}}}}

\newcommand{\Sched}{\ensuremath{\mathfrak{S}}\xspace}
\newcommand{\Scheds}{\ensuremath{\mathrm{Sched}}\xspace}
\newcommand{\SimpleScheds}{\ensuremath{\mathrm{SSched}}\xspace}
\newcommand{\FinitePaths}{\ensuremath{\mathrm{Paths}_\mathrm{fin}}}
\newcommand{\Paths}{\ensuremath{\mathrm{Paths}}}

\newcommand{\last}[1]{\ensuremath{\mathrm{last}(#1)}\xspace}
\newcommand{\probm}[2]{\ensuremath{\mathcal{P}^{#1}_{#2}}\xspace}
\newcommand{\expm}[2]{\ensuremath{\mathcal{E}^{#1}_{#2}}\xspace}
\newcommand{\probmax}[1]{\ensuremath{\mathcal{P}^{\mathrm{max}}_{#1}}\xspace}
\newcommand{\expmax}[1]{\ensuremath{\mathcal{E}^{\mathrm{max}}_{#1}}\xspace}
\newcommand{\probmin}[1]{\ensuremath{\mathcal{P}^{\mathrm{min}}_{#1}}\xspace}
\newcommand{\expmin}[1]{\ensuremath{\mathcal{E}^{\mathrm{min}}_{#1}}\xspace}
\newcommand{\probopt}[1]{\ensuremath{\mathcal{P}^{\mathit{opt}}_{#1}}\xspace}
\newcommand{\expopt}[1]{\ensuremath{\mathcal{E}^{\mathit{opt}}_{#1}}\xspace}
\newcommand{\Rew}{\ensuremath{\mathit{Rew}}\xspace}
\newcommand{\Rewe}{\ensuremath{\mathit{Rew}_\mathrm{exp}}\xspace}
\newcommand{\Rewb}{\ensuremath{\mathit{Rew}_\mathrm{bound}}\xspace}
\newcommand{\reachprob}[2]{\ensuremath{\mathrm{P}_{\!{#1}}({#2})}}
\newcommand{\stepreachprob}[3]{\ensuremath{\mathrm{P}^{\mathrm{T}[\leq{#1}]}_{\!{#2}}({#3})}}
\newcommand{\rewardreachprob}[3]{\ensuremath{\mathrm{P}^{\mathrm{R}[\leq{#1}]}_{\!{#2}}({#3})}}
\newcommand{\pathreward}[1]{\ensuremath{\mathrm{reward}(#1)}\xspace}
\newcommand{\reachreward}[2]{\ensuremath{\mathrm{R}_{\!{#1}}({#2})}}
\newcommand{\stepreachreward}[3]{\ensuremath{\mathrm{R}^{\mathrm{T}[\leq{#1}]}_{\!{#2}}({#3})}}
\newcommand{\rewardreachreward}[3]{\ensuremath{\mathrm{R}^{\mathrm{R}[\leq{#1}]}_{\!{#2}}({#3})}}

\newcommand{\uplimit}[2]{\ensuremath{{#1}{\uparrow}_{#2}}\xspace}
\newcommand{\limit}[2]{\ensuremath{{#1}{\downarrow}_{#2}}\xspace}
\newcommand{\Dom}[1]{\ensuremath{\mathrm{Dom}({#1})}\xspace}
\newcommand{\unroll}{\mathrm{unfold}}

\iftrue
\newcommand{\ah}[1]{{\color{teal}\textbf{AH:} #1}}
\newcommand{\mh}[1]{{\color{blue}\textbf{MH:} #1}}
\else
\newcommand{\ah}[1]{}
\newcommand{\mh}[1]{}
\fi
\usepackage{color}

\begin{document}

\title{%
Efficient Algorithms for Time- and\\ Cost-Bounded Probabilistic Model Checking%
}
\author{
Ernst Moritz Hahn\inst{1} \and
Arnd Hartmanns\inst{2}
}
\institute{
Institute of Software, Chinese Academy of Sciences, Beijing, China \and
University of Twente, Enschede, The Netherlands
}
\date{\today}
\maketitle

\begin{abstract}
In the design of probabilistic timed systems, bounded requirements concerning behaviour that occurs within a given time, energy, or more generally cost budget are of central importance.
Traditionally, such requirements have been model-checked via a reduction to the unbounded case by unfolding the model according to the cost bound.
This exacerbates the state space explosion problem and significantly increases runtime.
In this paper, we present three new algorithms to model-check time- and cost-bounded properties for Markov decision processes and probabilistic timed automata that avoid unfolding.
They are based on a modified value iteration process, on an enumeration of schedulers, and on state elimination techniques.
We can now obtain results for \emph{any} cost bound on a single state space no larger than for the corresponding unbounded or expected-value property.
In particular, we can naturally compute the cumulative distribution function at no overhead.
We evaluate the applicability and compare the performance of our new algorithms and their implementation on a number of case studies from the literature.
\end{abstract}

\section{Introduction}
\label{sec:Introduction}

Markov decision processes (MDP,~\cite{Put94}) and probabilistic timed automata (PTA, \cite{KNSS02}) are two popular formal models for probabilistic (real-time) systems.
The former combine nondeterministic choices, which can be due to concurrency, unquantified uncertainty, or abstraction, with discrete probabilistic decisions, which represent quantified uncertainty \eg due to environmental influences or in randomised algorithms.
The latter additionally provide facilities to model hard real-time behaviour and constraints as in timed automata.
Given an MDP or a PTA, queries like ``what is the worst-case probability to reach an unsafe system state'' or ``what is the minimum expected time to termination'' can be answered via probabilistic model checking~\cite{BDFK16,NPS13}. 
Although limited by the state space explosion problem, it works well on relevant case studies.

In practice, an important class of queries relates to cost- or reward-bounded properties, such as ``what is the probability of a message to arrive after at most three transmission attempts'' in a communication protocol, or ``what is the expected energy consumption within the first five hours after waking from sleep'' in a battery-powered device.
Costs and rewards are the same concept, and we will prefer the term \emph{reward} in the remainder of this paper.
In the properties above, we have three different rewards:
retransmission attempts (accumulating reward 1 for each attempt), energy consumption (accumulating reward at a state-dependent wattage), and time (accumulating reward at a rate of~1 in all states).
To compute reward-bounded properties, the traditional approach for discrete-time probabilistic models is to \emph{unfold} the state space~\cite{AHK03}:
in addition to the current state of the model, one keeps track of the reward accumulated so far, up to the specified bound value $b \in \NN$ (\eg $b = 3$ in the first property above).
This blows up the size of the state space linearly in $b$, and often causes the model checking process to run out of memory.
The situation for PTA is no different:
the bounded case is reduced to the unbounded one by extending the model, \eg by adding a new clock variable that is never reset to check time-bounded properties~\cite{NPS13}, with the same effect on state space size.
Using a digital clocks semantics~\cite{KNPS06}, the analysis of PTA with rewards can be reduced to MDP model checking.

In this paper, we present three new algorithms for probabilistic model checking of reward-bounded properties \emph{without unfolding}.
The first algorithm is a modification of standard unbounded value iteration.
The other algorithms use different techniques---scheduler enumeration with either value iteration or Markov chain state elimination, and MDP state elimination---to reduce the model such that all remaining transitions correspond to accumulating a reward of~$1$.
A \emph{reward-bounded} property with bound~$b$ in the original model then corresponds to a \emph{step-bounded} property with bound $b$ in the reduced model.
We use standard step-bounded value iteration~\cite{Put94} to check these properties efficiently.

Common to all three algorithms is that there is no blowup in the number of states due to unfolding.
There no blowup at all in the first algorithm.
If we could previously check an unbounded property with a given amount of memory, we can now check the corresponding bounded property for \emph{any} bound value~$b$, too.
In fact, when asked for the probability to reach a certain set of states with accumulated reward~$\leq b$, all three algorithms actually compute the sequence of probabilities for bounds $0, \dots, b$.
At no overhead, we thus obtain the cumulative (sub)distribution function over the bound up to~$b$.
If it converges in a finite number of steps, then we can detect this step by keeping track of the maximum error in the value iterations.
We may thus obtain the entire function without the user having to specify an a priori bound.
From a practical perspective, this means that with memory previously sufficient to compute the (unbounded) expected reward (\ie the mean or first moment of the underlying distribution), we can now obtain the entire distribution (\ie all moments).
Domain experts may accept a mean as a good first indicator of a system's behaviour, but are ultimately more interested in the actual shape of the distribution as a whole.

We have implemented all three algorithms in the \mcsta tool, which is available as part of the \toolset~\cite{HH14}.
After describing the algorithms in \Cref{sec:Algorithms} and the implementation in \Cref{sec:Implementation}, we use a number of case studies from the literature to evaluate the applicability and performance in \Cref{sec:Experiments}.

\paragraph{Related work.}
Only very recently has a procedure to handle reward-bounded properties without unfolding been described for MDP~\cite{BDFK16,HHS16}.
It works by solving a sequence of linear programming problems.
Solution vectors need to be stored to solve the subsequent instances.
Linear programming does not scale to large MDP, and we are not currently aware of a publicly available implementation of this procedure.
Yet, the underlying idea is similar to the first algorithm that we present in this paper, which uses value iteration instead.
For the soft real-time model of Markov automata, which includes MDP as a special case, reward-bounded properties can be turned into time-bounded ones~\cite{HBWFHB15}.
However, this only works for rewards associated to Markovian states, whereas immediate states (\ie the MDP subset of Markov automata) always implicitly get zero reward.

\section{Preliminaries}
\label{sec:Preliminaries}

\NN is $\set{ 0, 1, \dots }$, the set of natural numbers.
\QQplus is the set of positive rational numbers.
\RRpluszero is $[0, \infty)$, the set of nonnegative real numbers.
$\powerset{S}$~denotes the powerset of~$S$.
$\Dom{f}$ is the domain of the function~$f$.

\begin{definition}
A (discrete) \emph{probability distribution} over a set~$\varOmega$ is a function $\mu \in \varOmega \to [0, 1]$ such that $\support{\mu} \defeq \set{\omega \in \varOmega \mid \mu(\omega) > 0}$ is countable and  $\sum_{\omega \in \support{\mu}}{\mu(\omega)} = 1$.
$\Dist{\varOmega}$ is the set of all probability distributions over~$\varOmega$.
$\Dirac{s}$ is the \emph{Dirac distribution}\index{Dirac} for~$s$, defined by $\Dirac{s}(s) = 1$.
\end{definition}

\subsubsection{Markov Decision Processes}

To move from one state to another in a Markov decision process, first a transition is chosen nondeterministically; each transition then leads into a probability distribution over successor states.

\begin{definition}
A \emph{Markov decision process}~(MDP) is a tuple
$M =\tuple{\States, \Alphabet, \Trans, \InitialState}$
where
$\States$ is a finite set of states,
$\Alphabet$ is a finite set of actions,
$\Trans\in \States \to \powerset{\Alphabet \times \Dist{\States}}$ is the transition function, and
$\InitialState \in \States$ is the initial state.
For all $s \in \States$, we require that $\Trans(s)$ is finite non-empty, and that if $\tuple{a, \mu} \in \Trans(s)$ and $\tuple{a, \mu'} \in \Trans(s)$ then $\mu = \mu'$.
We call $s \in \States$ \emph{deterministic} if $|\Trans(s)| = 1$.
\end{definition}
We write $s \xtr{a}_\Trans \mu$ for $\exists\,a, \mu \colon \tuple{a, \mu} \in \Trans(s)$ and call it a \emph{transition}.
We write $s \xtr{a}_\Trans s'$ if additionally $s' \in \support{\mu}$.
If $\Trans$ is clear from the context, we write $\xtr{}$ instead of $\xtr{}_\Trans$.
Graphically, we represent transitions as action-labelled lines to an intermediate node from which weighted \emph{branches} lead to successor states.
We may omit action labels out of deterministic states as well as the intermediate node and probability~$1$ for transitions into Dirac distributions.

\begin{definition}
A \emph{reward structure} for $M$ is a function $\Rew \in \States \times \Alphabet \times \States \to \RRpluszero$ such that $\Rew(\tuple{s, a, s'}) \neq 0 \implies s \xtr{a} s'$.
It associates a \emph{branch reward} to each choice of action and successor state for all transitions.
\end{definition}
\begin{figure}[t]
\begin{minipage}[b]{.4\textwidth}
\centering
\begin{tikzpicture}[on grid,auto]
  \node[state] (s) {$s$};
  \coordinate[left=0.3 of s.west] (start);
  \node[state] (t) [right=1.75 of s] {$t$};
  \node[state] (u) [below=1.0 of t] {$u$};
  \node[state] (v) [right=1.75 of t,accepting,thin] {$v$};
  \node[state] (w) [below=1.0 of v] {$w$};
  \node[dot] (ssu) [below right=0.75 and 0.7 of s] {};
  \node[dot] (ttv) [right=0.75 of t] {};
  \node[dot] (uvw) [right=0.75 of u] {};
  ;
  \path[-]
    (s) edge[bend left=15] node[pos=0.85,inner sep=1pt] {\texttt{b}} (ssu)
    (t) edge[bend right] node[swap,pos=0.39,inner sep=1.5pt] {\texttt{d}} (ttv)
    (u) edge[bend right] node[swap,pos=0.39,inner sep=1.5pt] {\texttt{e}} (uvw)
  ;
  \path[->]
    (start) edge node {} (s)
    (ssu) edge[bend right,swap,pos=0.7] node[inner sep=1pt] {$0.5$} (u)
    (ssu) edge[out=-120,in=-90,pos=0.3] node[overlay,inner sep=1pt,align=center] {$0.5$,\\$+1$\phantom{,}} (s)
    (s) edge[bend left=20] node[pos=0.375,inner sep=1.5pt] {\texttt{a}} (t)
    (t) edge[bend left=20] node[pos=0.3,inner sep=2.0pt] {\texttt{c}} (s)
    (ttv) edge[out=90,in=90] node[swap,pos=0.61,inner sep=1pt] {~~~$0.8$,\,$+1$} (t)
    (ttv) edge[bend left,pos=0.6] node[inner sep=2pt] {$0.2$} (v)
    (uvw) edge[bend left] node[pos=0.1,inner sep=0.5pt] {$0.5$} (v)
    (uvw) edge[bend left,pos=0.6] node[swap,inner sep=3pt] {$0.5$} (w)
    (v) edge[loop,out=-30,in=30,looseness=4,pos=0.5] (v)
    (w) edge[loop,out=-30,in=30,looseness=4,pos=0.5] (w)
  ;
\end{tikzpicture}\vspace{0.5pt}
\caption{Example MDP $M^e$}
\label{fig:ExampleMDP}
\end{minipage}%
\begin{minipage}[b]{.6\textwidth}
\centering
\begin{tikzpicture}[on grid,auto]
  \node[state] (s) {$s$};
  \coordinate[left=0.3 of s.west] (start);
  \node[lstate] (snew) [below=1.125 of s] {$\strut s_\mathrm{new}$};
  \node[state] (t) [right=1.75 of s] {$t$};
  \node[state] (u) [below=1.0 of t] {$u$};
  \node[lstate] (tnew) [above right=0.5 and 1.75 of t] {$\strut t_\mathrm{new}$};
  \node[state] (v) [below right=0.375 and 1.75 of t,accepting,thin] {$v$};
  \node[lstate] (vnew) [right=1.0 of v] {$\strut v_\mathrm{new}$};
  \node[state] (w) [below=0.75 of v] {$w$};
  \node[dot] (ssu) [below right=0.75 and 0.7 of s] {};
  \node[dot] (ttv) [right=0.75 of t] {};
  \node[dot] (uvw) [right=0.75 of u] {};
  ;
  \path[-]
    (s) edge[bend left=15] node[pos=0.85,inner sep=1pt] {\texttt{b}} (ssu)
    (t) edge[bend right] node[pos=0.67,inner sep=1.2pt] {\texttt{d}} (ttv)
    (u) edge[bend right] node[swap,pos=0.4,inner sep=1.5pt] {\texttt{e}} (uvw)
  ;
  \path[->]
    (start) edge node {} (s)
    (ssu) edge[bend right,pos=0.35] node[inner sep=1pt] {$0.5$} (u)
    (ssu) edge[bend left] node[inner sep=0.5pt,pos=0.7] {$0.5$} (snew)
    (s) edge[bend left=20] node[pos=0.3,inner sep=1.5pt] {\texttt{a}} (t)
    (t) edge[bend left=20] node[pos=0.3,inner sep=2.0pt] {\texttt{c}} (s)
    (ttv) edge[bend left] node[pos=0.61,inner sep=0.5pt] {~~~$0.8$} (tnew)
    (ttv) edge[bend left,pos=0.7] node[inner sep=0.5pt] {$0.2$} (v)
    (uvw) edge[bend left] node[pos=0.25,inner sep=0.5pt] {$0.5$} (v)
    (uvw) edge[bend left,pos=0.8] node[swap,inner sep=1.5pt] {$0.5$} (w)
    (v) edge[pos=0.6] (vnew)
    (vnew) edge[loop,out=-20,in=20,looseness=3,pos=0.5] (vnew)
    (w) edge[overlay,loop,out=-30,in=30,looseness=4,pos=0.5] (w)
    (snew) edge[loop,out=160,in=200,looseness=3,pos=0.5] (snew)
    (tnew) edge[overlay,loop,out=-20,in=20,looseness=3,pos=0.5] (tnew)
  ;
\end{tikzpicture}
\caption{Transformed MDP $\limit{\limit{M^e}{F^e}}{\Rew^e}$}
\label{fig:ExampleMDPRestricted}
\end{minipage}
\end{figure}
\Cref{fig:ExampleMDP} shows an example MDP~$M^e$ with 5~states, 6~transitions and 10~branches.
States $u$, $v$ and $w$ are deterministic.
$M^e$ has one reward structure $\Rew^e$ with $\Rew^e(\tuple{s, \texttt{b}, s}) = \Rew^e(\tuple{t, \texttt{d}, t}) = 1$ and $\Rew^e(\mathit{tr}) = 0$ otherwise.

Using MDP directly to build complex models is cumbersome.
Instead, high-level formalisms such as \prism's~\cite{KNP11} guarded command language are used.
Aside from a parallel composition operator, they extend MDP with variables over finite domains that can be used in expressions to \eg enable/disable transitions.
The semantics of such a high-level model is an MDP whose states are the valuations of the variables.
This allows to compactly describe very large MDP.

\paragraph{Paths and schedulers.}
The semantics of an MDP $M = \tuple{\States, \Alphabet, \Trans, \InitialState}$ is captured by the notion of paths.
A path represents a concrete resolution of both nondeterministic and probabilistic choices:
\begin{definition}
A \emph{finite path} from $s_0$ to~$s_n$ of length~$n \in \NN$ is a finite sequence
$\pi_\mathrm{fin} = s_0\, a_0\, s_1\, a_1\, s_2 \dots a_{n-1}\, s_n$
where $s_i \in \States$ for all $i \in \set{ 0, \dots, n }$ and $s_i \xtr{a_i} s_{i+1}$ for all $i \in \set{ 0, \dots, n - 1 }$.
Let $|\pi_\mathrm{fin}| \defeq n$ and $\last{\pi_\mathrm{fin}} \defeq s_n$.
Given a reward structure \Rew, we define $\pathreward{\Rew, \pi_\mathrm{fin}} = \sum_{i=0}^{n-1} \Rew(\tuple{s_i, a_i, s_{i+1}})$.
$\FinitePaths(M)$ is the set of all finite paths from $\InitialState$.
An (infinite) \emph{path} starting from $s_0$ is an infinite sequence
$\pi = s_0\, a_0\, s_1\, a_1\, s_2 \dots$
where for all $i \in \NN$, we have that $s_i \in \States$ and $s_i \xtr{a_i} s_{i+1}$.
$\Paths(M)$ is the set of all infinite paths starting from $\InitialState$.
Given a set $F \subseteq \States$, let $\pi^{\diamond F}$ be the shortest prefix of $\pi$ that contains a state in~$F$, or $\bot$ if such a prefix does not exist.
\end{definition}
In contrast to a path, a scheduler (or \emph{adversary}, \emph{policy} or \emph{strategy}) only resolves the nondeterministic choices of~$M$:
\begin{definition}
\label{def:MDPReductionFunction}
A \emph{scheduler} is a function $\Sched \in \FinitePaths(M) \to \Dist{\Alphabet}$ such that $\forall\,s \in \States \colon a \in \support{\Sched(s)} \implies s \xtr{a} \mu$.
$\Scheds(M)$ is the set of all schedulers of~$M$.
\Sched~is \emph{reward-positional} for a reward structure \Rew if $\last{\pi_1} = \last{\pi_2}$ and  $\pathreward{\Rew, \pi_1} = \pathreward{\Rew, \pi_2}$ implies $\Sched(\pi_1) = \Sched(\pi_2)$, \emph{positional} if $\last{\pi_1} = \last{\pi_2}$ alone implies $\Sched(\pi_1) = \Sched(\pi_2)$, and \emph{deterministic} if $|\support{\Sched(\pi)}| = 1$, for all finite paths~$\pi$, $\pi_1$ and $\pi_2$, respectively.
A \emph{simple} scheduler is positional and deterministic.
It can thus be seen as a function in $\States \to \Alphabet$.
The set of all simple~schedulers of $M$ is $\SimpleScheds(M)$.
\end{definition}
Let $\limit{M}{\Sched_\mathrm{s}} \defeq\tuple{\States, \Alphabet, \Trans', \InitialState}$ with $\Trans'(s) \defeq \set{ \tuple{a, \mu} \mid s \xtr{a} \mu \wedge \Sched_\mathrm{s}(s) = a }$ for $\Sched_\mathrm{s} \in \SimpleScheds(M)$.
All states in $\limit{M}{\Sched_\mathrm{s}}$ are deterministic, \ie $\limit{M}{\Sched_\mathrm{s}}$ is a discrete-time Markov chain (DTMC).
Using the standard cylinder set construction~\cite{BK08}, a scheduler~$\Sched$ induces a probability measure $\probm{\Sched}{M}$ on measurable sets of paths starting from $\InitialState$.
Let $\expm{\Sched}{M}(\mathit{rv})$ denote the expectation of random variable $\mathit{rv}$ under this probability measure.
We define the \emph{extremal} values $\probmax{M}(\Pi) = \sup_{\Sched \in \Scheds(M)}\probm{\Sched}{M}(\Pi)$ and $\probmin{M}(\Pi) = \inf_{\Sched \in \Scheds(M)}\probm{\Sched}{M}(\Pi)$.
For expectations, $\expmax{M}$ and $\expmin{M}$ are defined analogously.

\paragraph{Properties.}
\label{sec:MDPProperties}
For an MDP $M$, reward structures $\Rewe$ and $\Rewb$, and a set of goal states $F \subseteq \States$, we define the following values for $\mathit{opt} \in \set{ \mathrm{max}, \mathrm{min} }$:
\begin{itemize}
\item
$\reachprob{\mathit{opt}}{F}$ is
the extremal probability of eventually reaching $F$, defined as $\probopt{M}(\Pi)$ where $\Pi$ is the set of paths in $\Paths(M)$ that contain a state in~$F$.
\item
$\stepreachprob{n}{\mathit{opt}}{F}$ is
the extremal probability of reaching $F$ via at most $n \in \NN$ transitions, defined as $\probopt{M}(\Pi^\mathrm{T}_n)$ where $\Pi^\mathrm{T}_n$ is the set of paths that have a prefix of length at most $n$ that contains a state in~$F$.
\item
$\rewardreachprob{n}{\mathit{opt}}{F}$ is
the extremal probability of reaching $F$ with accumulated reward $\Rewb$ at most~$n \in \RR$, defined as $\probopt{M}(\Pi^\mathrm{R}_n)$ where $\Pi^\mathrm{R}_n$ is the set of paths that have a prefix~$\pi_\mathrm{fin}$ containing a state in~$F$ with $\pathreward{\Rewb, \pi_\mathrm{fin}} \leq n$.
\item
$\reachreward{\mathit{opt}}{F}$ is
the expected accumulated reward \Rewe when reaching a state in $F$, defined as $\expopt{M}(f(F))$ where $f(F)(\pi) \defeq \pathreward{\Rewe, \pi^{\diamond F}}$ for $\pi^{\diamond F} \neq \bot$ and $f(F)(\bot) \defeq \infty$,
\item
$\stepreachreward{n}{\mathit{opt}}{F}$ is
the expected accumulated reward \Rewe when reaching a state in $F$ via at most $n \in \NN$ transitions, defined as $\expopt{M}(f^T_n(F))$ where $f^T_n(F)(\pi) \defeq \pathreward{\Rewe, \pi^{\diamond F}}$ if $\pi^{\diamond F} \neq \bot \wedge |\pi^{\diamond F}| \leq n$ and $f(F)(\pi) \defeq \infty$ otherwise.
\item
$\rewardreachreward{n}{\mathit{opt}}{F}$ is
the expected accumulated reward \Rewe when reaching $F$ with accumulated reward \Rewb at most $n \in \RR$, defined as $\expopt{M}(f^R_n(F))$ where $f^R_n(F)(\pi) \defeq \pathreward{\Rewe, \pi^{\diamond F}}$ if $\pi^{\diamond F} \neq \bot \wedge \pathreward{\Rewb, \pi^{\diamond F}} \leq n$ and $f(F)(\pi) \defeq \infty$ otherwise.
\end{itemize}
We refer to these values as \emph{unbounded}, \emph{step-bounded} or \emph{reward-bounded} reachability probabilities and expected accumulated rewards, respectively.

\begin{algorithm}[t]
\Fn{$\texttt{UnboundedVI}(V, \tuple{\States, \Trans, \InitialState}, \mathit{opt} \in \set{ \max, \min })$}{
  \Repeat{$\mathit{error} < \epsilon$}{
    $\mathit{error} := 0$\;
    \ForEach{$s \in \States$}{
      $v_\mathit{new} := \mathit{opt}\,\set{ \sum_{s' \in \support{\mu}}{\mu(s') \cdot V(s') \mid s \xtr{a} \mu}}$\;
      \lIf{$v_\mathit{new} > 0$}{
        $\mathit{error} := \max\,\set{\mathit{error}, |v_\mathit{new} - V(s)| / v_\mathit{new} }$
      }
      $V(s) := v_\mathit{new}$
    }
  }
}
\caption{\small Value iteration for unbounded reachability}
\label{alg:ValueIterationUnbounded}
\end{algorithm}

\begin{theorem}
\label{theo:SchedulerPower}
For an unbounded property, there exists an optimal simple scheduler, \ie one that attains the extremal value~\cite{BK08}.
For a reward-bounded property, there exists an optimal deterministic reward-positional scheduler for \Rewb~\cite{HHS16}.
\end{theorem}
Continuing our example, let $\Rewb = \Rew^e$ and $F^e = \set{ v }$.
We maximise the probability to eventually reach $F^e$ in $M^e$ by always scheduling \texttt{a} in~$s$ and \texttt{d} in $t$, so $\reachprob{\mathrm{max}}{F^e} = 1$ with a simple scheduler.
We get $\rewardreachprob{0}{\mathrm{max}}{F^e} = 0.25$ by scheduling \texttt{b} in~$s$.
For higher bound values, simple schedulers are no longer sufficient:
we get $\rewardreachprob{1}{\mathrm{max}}{F^e} = 0.4$ by first trying \texttt{a} then \texttt{d}, but falling back to \texttt{c} then \texttt{b} if we return to~$t$.
We maximise the probability for higher bound values $n$ by trying \texttt{d} until the accumulated reward is $n - 1$ and then falling back to~\texttt{b}.

\paragraph{Model checking.}
\label{sec:MDPModelChecking}
Probabilistic model checking works in two phases:
(1)~state space \emph{exploration} turns a given high-level model into an in-memory representation of the underlying MDP, then
(2)~a numerical \emph{analysis} computes the value of the property of interest.
In phase~1, the goal states are made absorbing:

\begin{algorithm}[t]
\Fn{$\texttt{StepBoundedVI}(V, \tuple{\States, \Trans, \InitialState}, n, \mathit{opt} \in \set{ \max, \min })$}{
  \For{$i = 1$ \textbf{to} $n$}{
    $V_\mathrm{old} := V$\;
    \ForEach{$s \in \States$}{
      $V(s) := \mathit{opt}\,\set{ \sum_{s' \in \support{\mu}}{\mu(s') \cdot V_\mathrm{old}(s') \mid s \xtr{a} \mu} }$\;
    }
  }
}
\caption{\small Value iteration for step-bounded reachability}
\label{alg:ValueIterationStepBounded}
\end{algorithm}

\begin{definition}
Given $M = \tuple{\States, \Alphabet, \Trans, \InitialState}$ and $F \subseteq \States$, we define the \emph{$F$-absorbing} MDP as $\limit{M}{F} = \tuple{\States, \Alphabet \cup \set{ \tau }, \Trans', \InitialState}$ with $\Trans'(s) = \set{\tuple{\tau, \Dirac{s}}}$ for all $s \in F$ and $\Trans'(s) = \Trans(s)$ otherwise.
We set $\Rewb(\tuple{s, \tau, s}) = 1$ for all $s \in F$.
For $s \in \States$, we define $M[s] = \tuple{\States, \Alphabet, \Trans, s}$ as the MDP with initial state changed to~$s$.
\end{definition}
An efficient algorithm for phase~2 is value iteration, which iteratively improves a \emph{value vector} containing for each state an approximation of the property's value.
The value iteration procedure for unbounded reachability probabilities is shown as~\Cref{alg:ValueIterationUnbounded}, while the one for step-bounded reachability with bound $n$ is shown as \Cref{alg:ValueIterationStepBounded}.
Let $V = \set{ s \mapsto 1 \mid s \in F} \cup \set{ s \mapsto 0 \mid s \in \States \setminus F }$ initially.
Then after termination of $\texttt{UnboundedVI}(V, \limit{M}{F}, \mathit{opt})$, we have $V(s) = \reachprob{\mathit{opt}}{F}$ in $M[s]$ for all $s \in \States$.
After termination of $\texttt{StepBoundedVI}(V_0, \limit{M}{F}, n, \mathit{opt})$, we would instead have $V(s) = \stepreachprob{n}{\mathit{opt}}{F}$ in $M[s]$ for all $s \in \States$.
The algorithms for expected rewards are very similar throughout~\cite{Put94}.

The traditional way to model-check reward-bounded properties is to \emph{unfold} the model according to the accumulated reward:
a reward structure is turned into a variable $v$ in the model prior to phase~1, with transition reward $r$ corresponding to an assignment~$v := v + r$.
To check $\rewardreachprob{n}{\mathit{opt}}{F}$, phase~1 thus creates an MDP that is up to $n$ times as large as without unfolding.
In phase~2, $\reachprob{\mathit{opt}}{F'}$ is checked where $F'$ corresponds to the states in $F$ where additionally $v \leq n$ holds.

\subsubsection{Probabilistic Timed Automata}

Probabilistic timed automata (PTA~\cite{KNSS02}) extend MDP with \emph{clocks} and \emph{clock constraints} as in timed automata to model real-time behaviour and requirements.
A reward structure for a PTA defines two kinds of rewards:
\emph{edge rewards} are accumulated when an action is performed as in MDP, and \emph{rate rewards} accumulate at a certain rate over time.
Time itself is a special rate reward that is always~1.

There are currently three approaches to model-check PTA~\cite{NPS13}.
Of these, only the digital clocks approach~\cite{KNPS06} preserves expected rewards.
It works by replacing the clock variables by bounded integers and adding self-loop edges to increment them synchronously as long as time can pass.
The reward of each of these self-loop edges is the current rate reward.
The result is (a high-level model of) a finite \emph{digital clocks MDP}.
All the algorithms that we develop for MDP in this paper can thus be applied to PTA as well, with one restriction:
general rate-reward-bounded properties are undecidable~\cite{BCJ09}.
We summarise the decidability of the different kinds of reward-bounded properties for PTA in \Cref{tab:PTADecidability}.

\begin{table}[t]
\caption{Decidability of PTA properties}
\label{tab:PTADecidability}
\centering
\setlength{\tabcolsep}{5pt}
\begin{tabular}{c|c|ccc}
& & \multicolumn{3}{c}{bounded by} \\
& unbounded & edge rewards & time & rate rewards \\\hline
reachability probabilities & \cmark & \cmark~\cite{KNPS06} & \cmark~\cite{KNPS06} & \xmark~\cite{BCJ09} \\
expected accumulated\ rewards & \cmark & \cmark~\cite{KNPS06} & \cmark~\cite{KNPS06} & \xmark\phantom{~\cite{BCJ09}} \\
\end{tabular}
\end{table}

\section{Algorithms}
\label{sec:Algorithms}

We now present three new algorithms that allow the computation of reward-bounded reachability probabilities and expected accumulated rewards on MDP without unfolding.
In essence, they all emulate (deterministic) reward-positional schedulers in different ways.
For clarity, the algorithms as we present them here compute reachability probabilities.
They can easily be changed to compute expected rewards by additionally keeping track of rewards as values are updated or transitions are merged.
We also assume that the reward structure \Rewb only takes values zero or one, \ie $\Rewb(t) \in \set{0, 1}$ for all $t \in \States \times \Alphabet \times \States$, and that the property bound $n$ is in $\NN$.
This is without loss of generality in practice:
If $\Rewb(t) = r \in \NN$ with $r > 1$ for some $t$, then we can replace it by a chain of $r$ transitions with reward~$1$.
If there is a value in $\QQplus \setminus \NN$, then we need to find the least common multiple $d_\mathrm{lcm} \in \NN$ of the denominators of all the values, multiply them (and the bound) by $d_\mathrm{lcm}$, and proceed as in the integer case.

For all three algorithms, we need two transformations that redirect the reward-one branches.
The first one, $\uplimit{}{\Rewb}$, redirects each such branch to an absorbing copy $s_\mathrm{new}$ of the transition's origin state~$s$, while the second one, $\limit{}{\Rewb}$, redirects to a copy $s''_\mathrm{new}$ of the target state~$s''$ of the branch:

\begin{definition}
Given $M = \tuple{\States, \Alphabet, \Trans, \InitialState}$ and a reward structure \Rewb, we define $\uplimit{M}{\Rewb}$ as $\tuple{\States \uplus \States_\mathrm{new}, \Alphabet \cup \set{ \tau }, \Trans^\uparrow, \InitialState}$ and $\limit{M}{\Rewb}$ as $\tuple{\States \uplus \States_\mathrm{new}, \Alphabet \cup \set{ \tau }, \Trans^\downarrow, \InitialState}$ with $\States_\mathrm{new} = \set{ s_\mathrm{new} \mid s \in \States }$,
\[
\Trans^x(s) = \begin{cases}
\set{ \tuple{a, \mathit{Conv}^x(s, a, \mu)} \mid \tuple{a, \mu} \in \Trans(s) } & \text{if } s \in \States\\
\set{ \tuple{\tau, \Dirac{s}} } & \text{if } s \in \States_\mathrm{new}
\end{cases}
\]
and\vspace{-\baselineskip}
\[
\mathit{Conv}^x(s, a, \mu)(s') = \begin{cases}
\mu(s') & \text{if }\Rewb(\tuple{s, a, s'}) = 0\\
\mu(s'') & \text{if } x = {\uparrow} \text{, } s' = s_\mathrm{new} \text{ and }\Rewb(\tuple{s, a, s''}) = 1\\
\mu(s'') & \text{if } x = {\downarrow} \text{, } s' = s''_\mathrm{new} \text{ and }\Rewb(\tuple{s, a, s''}) = 1\\
0 & \text{otherwise.}
\end{cases}
\]
\end{definition}
For our example MDP $M^e$ and $F^e = \set{ v }$, we show $\limit{\limit{M^e}{F^e}}{\Rew^e}$ in \Cref{fig:ExampleMDPRestricted}.
All of our algorithms take a value vector $V$ as input, which they update.
$V$ must initially contain the probabilities to reach a goal state in $F$ with zero reward.
These can be computed via a call $\texttt{UnboundedVI}(V = V_F^0, \uplimit{\limit{M}{F}}{\Rewb}, \mathit{opt})$~with\\[1mm]
\centerline{$V_F^0 \defeq \set{ s \mapsto 1, s_\mathrm{new} \mapsto 1 \mid s \in F} \cup \set{ s \mapsto 0, s_\mathrm{new} \mapsto 0 \mid s \in \States \setminus F }$.}\\[1mm]
This is the only place where the $\uparrow$ transformation is needed.

\subsection{Modified Value Iteration}
\label{sec:ModifiedVI}

\begin{algorithm}[t]
\Fn{$\texttt{RewardBoundedIter}(V, M = \tuple{\States, \Alphabet, \Trans, \InitialState}, F, \Rewb, n, \mathit{opt})$}{
  \For{$i = 1$ \textbf{to} $n$}{
    \lForEach{$s_\mathrm{new} \in \States_\mathrm{new}$}{$V(s_\mathrm{new}) := V(s)$\label{alg:ModifiedValueIteration:Copy}}
    $\texttt{UnboundedVI}(V, \limit{\limit{M}{F}}{\Rewb}, \mathit{opt})$\label{alg:ModifiedValueIteration:Iterate}\;
  }
}
\caption{\small Modified value iteration for reward-bounded reachability}
\label{alg:ModifiedValueIteration}
\end{algorithm}

\begin{table}[t]
\begin{minipage}[t]{0.6\textwidth}
\caption{Running \Cref{alg:ModifiedValueIteration} on $M^e$}
\label{tab:ExampleModifiedVI}
\centering
\renewcommand{\arraystretch}{1.0}
\setlength{\tabcolsep}{2.3pt}
\begin{tabular}{r|cccccccc}
\multicolumn{1}{c|}{} & $s$ & $s_\mathrm{new}$ & $t$ & $t_\mathrm{new}$ & $u$ & $v$ & $v_\mathrm{new}$ & $w$ \\\hline
initially & 0.25 & 0 & 0.25 & 0 & 0.5 & 1 & 0 & 0 \\
copy (l.~\ref{alg:ModifiedValueIteration:Copy}) & 0.25 & 0.25 & 0.25 & 0.25 & 0.5 & 1 & 1 & 0 \\
iter (l.~\ref{alg:ModifiedValueIteration:Iterate}) & 0.40 & 0.25 & 0.40 & 0.25 & 0.5 & 1 & 1 & 0 \\
\end{tabular}
\end{minipage}%
\begin{minipage}[t]{0.4\textwidth}
\caption{\Cref{alg:StateElimination} on $M^e$}
\label{tab:ExampleEliminatedVI}
\centering
\renewcommand{\arraystretch}{1.0}
\setlength{\tabcolsep}{2.3pt}
\setlength{\fboxsep}{1.25pt}
\begin{tabular}{c|cccc}
\multicolumn{1}{c|}{} & $s$ & $t$ & $v$ & $w$ \\\hline
init & 0.25 & 0.25 & ~1~ & ~0~ \\
step & 0.4 & 0.4 & 1 & 0 \\
step & 0.52 & 0.52 & 1 & 0 \\
\end{tabular}
\end{minipage}
\end{table}

Our first new algorithm \alg{modvi} is shown as \Cref{alg:ModifiedValueIteration}.
It is a combination of the two value iteration algorithms for unbounded and step-bounded properties (\Cref{alg:ValueIterationUnbounded,alg:ValueIterationStepBounded}).
Where the unbounded one essentially finds an optimal simple scheduler and the step-bounded one simulates an optimal deterministic scheduler, our algorithm simulates reward-positional schedulers.
In a loop, it copies the results of the previous iteration for each regular state in $\States$ to the corresponding post-reward new state in $\States_\mathrm{new}$ (line~\ref{alg:ModifiedValueIteration:Copy}), then performs an unbounded value iteration (line~\ref{alg:ModifiedValueIteration:Iterate}) to compute new values.
Initially, $V$ contains the probabilities to reach a goal state with zero reward.
Copying the values from regular to new states corresponds to allowing $1$ more reward to be accumulated in the subsequent value iteration.
One run of the algorithm effectively computes a sequence of $n$ simple schedulers, which combined represent the optimal reward-positional scheduler.
After each loop iteration, $V(s)$ is $\rewardreachprob{i}{\mathit{opt}}{F}$ for $M[s]$.
The algorithm thus implicitly computes the probabilities for all bounds $i\leq n$; our implementation actually returns all of them explicitly.
\Cref{tab:ExampleModifiedVI} shows the evolution of the value vector as the algorithm proceeds over its first iterations on~$M^e$.

\subsection{Scheduler Enumeration}
\label{sec:AlgorithmsSched}

\begin{algorithm}[t]
\Fn{$\texttt{RewardBoundedSched}(V, M = \tuple{\States, \Alphabet, \Trans, \InitialState}, F, \Rewb, n, \mathit{opt})$}{
  $\Trans'' := \varnothing$, $M' := \limit{\limit{M}{F}}{\Rewb} = \tuple{\States \uplus \States_\mathrm{new}, \Alphabet', \Trans', \dots}$\;
  \ForEach{$s \in \set{ \InitialState } \cup \set{ s'' \mid \exists\, s' \colon s' \xtr{a}_{\Trans'} s'' \wedge \Rewb(\tuple{s', a, s''}) = 1 }$\label{alg:SchedulerEnumeration:OuterLoop}}{
    \ForEach(\tcp*[f]{simple scheduler enumeration}){$\Sched \in \SimpleScheds(M'[s])$\label{alg:SchedulerEnumeration:InnerLoop}}{
      $\Trans''(s) := \Trans''(s) \cup \set{ \texttt{ComputeProbs}(\limit{M'[s]}{\Sched}) }$\label{alg:SchedulerEnumeration:BuildTrans}\;
    }
  }
  $\Trans'' := \Trans'' \cup \set{ \bot \mapsto \set{ \tuple{\tau, \Dirac{\bot}} } }$, $V(\bot) := 0$\;
  $\texttt{StepBoundedVI}(V, \tuple{\Dom{T''}, \Alphabet', T'', \InitialState}, n, \mathit{opt})$\label{alg:SchedulerEnumeration:VI}\tcp*{step-bounded iteration}
}
\Fn{$\texttt{ComputeProbs}(M = \tuple{\States \uplus \States_\mathrm{new}, \ldots}\text{ deterministic})$}{
  $\mu := \set{ s \mapsto \reachprob{\mathit{opt}}{\set{ s_\mathrm{new} }} \mid s_\mathrm{new} \in \States_\mathrm{new} }$\;
  \Return{$\mu \cup \set{ \bot \mapsto 1 - \sum_{s_\mathrm{new} \in \States_\mathrm{new}}{\mu(s)} }$}\;
}
\caption{\small Reward-bounded reachability via scheduler enumeration}
\label{alg:SchedulerEnumeration}
\end{algorithm}

Our next algorithm, \alg{senum}, is summarised as \Cref{alg:SchedulerEnumeration}.
The idea of \texttt{Reward\-Bounded\-Sched} is to replace the entire sub-MDP between a ``relevant'' state and the new states (that follow immediately after a reward-one branch) by \emph{one direct} transition to a distribution over the new states \emph{for each} simple scheduler.
The actual reward-bounded probabilities can be computed on the result MDP $M''$ using the standard step-bounded algorithm (line~\ref{alg:SchedulerEnumeration:VI}), since one \emph{step} now corresponds to a \emph{reward} of~$1$.
The simple schedulers are preserved by the model transformation, and the step-bounded value iteration combines them into reward-positional ones.
The latter implicitly computes the probabilities for all bounds $i \leq n$, and our implementation again returns all of them~explicitly.

The relevant states, which remain in the result MDP~$M''$, are the initial state plus those states~$s \in \States$ that have an incoming reward-one branch.
We iterate over them in line~\ref{alg:SchedulerEnumeration:OuterLoop}.
In an inner loop (line~\ref{alg:SchedulerEnumeration:InnerLoop}), we iterate over the simple schedulers for each relevant state.
For each scheduler, \texttt{ComputeProbs} determines the probability distribution $\mu$ over reaching each of the new states (accumulating $1$ reward on the way) or getting stuck in an end component without being able to accumulate any more reward ever (as $\mu(\bot)$).
A transition to preserve $\mu$ in the result MDP is created in line~\ref{alg:SchedulerEnumeration:BuildTrans}.
The total number of simple schedulers for $n$ states with max.\ fan-out $m$ is in $\mathcal{O}(m^n)$, but we expect the number of schedulers that actually lead to different distributions from one relevant state up to the next reward-one steps to remain manageable.
The efficiency of the implementation hinges on a good procedure to enumerate all of but no more than the necessary schedulers.

\texttt{ComputeProbs} can be implemented in two ways:
either using standard unbounded value iteration, once for $p_F$ and once for each new state, or---since $\limit{M'[s]}{\Sched}$ is a DTMC---using DTMC \emph{state elimination}~\cite{HHZ11}.
The latter successively eliminates the non-new states as shown schematically in \Cref{fig:DTMCStateElimination} while preserving the reachability probabilities, all in one go.

\subsection{State Elimination}

\begin{figure}[t]
\begin{minipage}[b]{.5\textwidth}
\centering
\begin{tikzpicture}[on grid,auto]
  \node[state] (s) {$s$};
  \coordinate[above right=0.33 and 0.33 of s] (off1);
  \coordinate[above right=0.6 and 0.8 of s] (off2);
  \node[state] (t) [right=2 of s] {$t$};
  \node[state] (u1) [above right=0.5 and 2 of t] {$u_1$};
  \node[] (udots) [right=1.7 of t] {$\rvdots$};
  \node[state] (un) [below right=0.5 and 2 of t] {$u_n$};
  \node[] (to) [below=0.85 of t] {$\Downarrow$};
  \node[state] (sp) [below=2 of s] {$s$};
  \coordinate[above right=0.33 and 0.33 of sp] (off1p);
  \coordinate[above right=0.6 and 0.8 of sp] (off2p);
  \node[state] (u1p) [above right=0.5 and 4 of sp] {$u_1$};
  \node[] (udotsp) [right=3.7 of sp] {$\rvdots$};
  \node[state] (unp) [below right=0.5 and 4 of sp] {$u_n$};
  ;
  \path[-]
    (s) edge[] node {} (off1)
    (off1) edge[densely dashed,bend left=17.5] node[pos=0.25,inner sep=0.5pt] {$p_d$} (off2)
    (sp) edge[] node {} (off1p)
    (off1p) edge[densely dashed,bend left=17.5] node[pos=0.25,inner sep=0.5pt] {$p_d$} (off2p)
  ;
  \path[->]
    (s) edge[] node[pos=0.5,inner sep=1.5] {$p_a$} (t)
    (t) edge[loop,out=60,in=120,looseness=4] node[swap,pos=0.5,inner sep=1.5pt] {$p_c$} (t)
    (t) edge[bend right=10] node[pos=0.5,inner sep=1.5pt] {$p_{b_1}$} (u1)
    (t) edge[bend left=10] node[swap,pos=0.5,inner sep=1.5pt] {$p_{b_n}$} (un)
    (sp) edge[bend right=5] node[pos=0.5,inner sep=1.5pt] {$p_a\!\cdot\!\frac{p_{b_1}}{1-p_c}$} (u1p)
    (sp) edge[bend left=5] node[swap,pos=0.5,inner sep=1.5pt] {$p_a\!\cdot\!\frac{p_{b_n}}{1-p_c}$} (unp)
  ;
\end{tikzpicture}\vspace{8.25pt}
\caption{DTMC state elimination~\cite{HHZ11}}
\label{fig:DTMCStateElimination}
\end{minipage}%
\begin{minipage}[b]{.5\textwidth}
\centering
\begin{tikzpicture}[on grid,auto]
  \node[state] (s) {$s$};
  \coordinate[above right=0.33 and 0.33 of s] (off1);
  \coordinate[above right=0.66 and 0.66 of s] (off2);
  \node[dot] (d1) [right=0.75 of s] {};
  \coordinate[above right=0.33 and 0.33 of d1] (offd1);
  \coordinate[above right=0.6 and 0.8 of d1] (offd2);
  \node[state] (t) [right=2 of s] {$t$};
  \node[dot] (d2) [above right=0.25 and 0.75 of t] {};
  \node[dot] (d3) [below right=0.25 and 0.75 of t] {};
  \node[lstate] (u11) [above right=0.75 and 2 of t,inner sep=1.5pt] {$\strut u_{11}$};
  \node[] (udots) [below right=0.015 and 0.6 of t] {\scalebox{.75}{$\rvdots$}};
  \node[] (u1dots) [above right=0.46 and 1.6 of t] {\scalebox{.5}{$\vdots$}};
  \node[lstate] (u1n) [above right=0.25 and 2 of t,inner sep=1.5pt] {$\strut u_{1n}$};
  \node[lstate] (un1) [below right=0.75 and 2 of t,inner sep=1.5pt] {$\strut u_{nn}$};
  \node[] (undots) [below right=0.375 and 1.6 of t] {\scalebox{.5}{$\vdots$}};
  \node[lstate] (unn) [below right=0.25 and 2 of t,inner sep=1.5pt] {$\strut u_{n1}$};
  \node[] (to) [below left=0.7 and 0.85 of t] {$\Downarrow$};
  \node[state] (sp) [below=2 of s] {$s$};
  \coordinate[above right=0.33 and 0.33 of sp] (off1p);
  \coordinate[above right=0.66 and 0.66 of sp] (off2p);
  \node[dot] (d2p) [above right=0.25 and 1.5 of sp] {};
  \coordinate[above right=0.33 and 0.33 of d2p] (offd2p);
  \coordinate[above right=0.6 and 0.8 of d2p] (offd2ap);
  \node[dot] (d3p) [below right=0.25 and 1.5 of sp] {};
  \coordinate[below right=0.33 and 0.33 of d3p] (offd3p);
  \coordinate[below right=0.6 and 0.8 of d3p] (offd3ap);
  \node[lstate] (u11p) [above right=0.75 and 4 of sp,inner sep=1.5pt] {$\strut u_{11}$};
  \node[] (udotsp) [below right=0.015 and 1.35 of sp] {\scalebox{.75}{$\rvdots$}};
  \node[] (u1dotsp) [above right=0.485 and 3.6 of sp] {\scalebox{.5}{$\rvdots$}};
  \node[lstate] (u1np) [above right=0.25 and 4 of sp,inner sep=1.5pt] {$\strut u_{1n}$};
  \node[lstate] (un1p) [below right=0.75 and 4 of sp,inner sep=1.5pt] {$\strut u_{nn}$};
  \node[] (undotsp) [below right=0.515 and 3.6 of sp] {\scalebox{.5}{$\rvdots$}};
  \node[lstate] (unnp) [below right=0.25 and 4 of sp,inner sep=1.5pt] {$\strut u_{n1}$};
  ;
  \path[-]
    (s) edge[] node {} (off1)
    (off1) edge[densely dashed] node[pos=0.25,inner sep=0.75pt] {\texttt{c}} (off2)
    (d1) edge[] node {} (offd1)
    (offd1) edge[densely dashed,bend left=17.5] node[pos=0.25,inner sep=0.5pt] {$p_d$} (offd2)
    (s) edge[] node[inner sep=1pt] {\texttt{a}} (d1)
    (t) edge[] node[inner sep=1pt,pos=0.85] {$\texttt{b}_1$} (d2)
    (t) edge[] node[swap,inner sep=0pt,pos=0.85] {$\texttt{b}_n$} (d3)
    (sp) edge[] node {} (off1p)
    (off1p) edge[densely dashed] node[pos=0.25,inner sep=0.75pt] {\texttt{c}} (off2p)
    (d2p) edge[bend right=4] node {} (offd2p)
    (offd2p) edge[densely dashed,bend left=17.5] node[pos=0.25,inner sep=0.5pt] {$p_d$} (offd2ap)
    (d3p) edge[bend left=4] node {} (offd3p)
    (offd3p) edge[densely dashed,bend right=17.5] node[swap,pos=0.25,inner sep=0.5pt] {$p_d$} (offd3ap)
    (sp) edge[bend right=5] node[inner sep=1pt,pos=0.85] {$\texttt{ab}_1$} (d2p)
    (sp) edge[bend left=5] node[swap,inner sep=0.25pt,pos=0.85] {$\texttt{ab}_n$} (d3p)
  ;
  \path[->]
    (d1) edge[] node[pos=0.5,inner sep=1.5pt] {$p_a$} (t)
    (d2) edge[out=45,in=90,looseness=1.5] node[swap,pos=0.5,inner sep=1.5pt] {$p_{c_1}~~~$} (t)
    (d3) edge[out=-45,in=-90,looseness=1.5] node[pos=0.5,inner sep=0.75pt] {$p_{c_n}~~~$} (t)
    (d2) edge[bend right=2] node[pos=0.75,inner sep=0.5pt] {$p_{b_{11}}$} (u11)
    (d2) edge[bend left=5] node[swap,pos=0.5,inner sep=1.5pt] {$p_{b_{1n}}$} (u1n)
    (d3) edge[bend left=2] node[swap,pos=0.75,inner sep=0.0pt] {$p_{b_{nn}}$} (un1)
    (d3) edge[bend right=5] node[pos=0.5,inner sep=1.5pt] {$p_{b_{n1}}$} (unn)
    (d2p) edge[bend left=5] node[pos=0.95,inner sep=0.5pt] {$p_a\cdot$\scalebox{.8}{$\frac{p_{b_{11}}}{1-p_{c_1}}$}} (u11p)
    (d2p) edge[bend left=5] node[swap,pos=0.0,inner sep=0pt] {$p_a\cdot$\scalebox{.8}{$\frac{p_{b_{1n}}}{1-p_{c_1}}$}} (u1np)
    (d3p) edge[bend right=5] node[swap,pos=0.95,inner sep=0.0pt] {$p_a\cdot$\scalebox{.8}{$\frac{p_{b_{nn}}}{1-p_{c_1}}$}} (un1p)
    (d3p) edge[bend right=5] node[pos=1,inner sep=0.5pt] {$p_a\cdot$\scalebox{.8}{$\frac{p_{b_{n1}}}{1-p_{c_1}}$}} (unnp)
  ;
\end{tikzpicture}
\caption{MDP state elimination}
\label{fig:MDPStateElimination}
\end{minipage}
\end{figure}

Instead of a probability-preserving DTMC state elimination for each scheduler as in \alg{senum}, our third algorithm \alg{elim} performs a new scheduler- and probability-preserving state elimination on the entire MDP as shown schematically in \Cref{fig:MDPStateElimination}.
Observe how this elimination process preserves the options that simple schedulers have, and in particular relies on their positional character to be able to redistribute the loop probabilities $p_{c_i}$ onto the same transition only.

\begin{algorithm}[t]
\Fn{$\texttt{RewardBoundedElim}(V, M = \tuple{\States, \Alphabet, \Trans, \InitialState}, F, \Rewb, n, \mathit{opt})$}{
  $M' := \limit{\limit{M}{F}}{\Rewb} = \tuple{\States \uplus \States_\mathrm{new}, \dots}$\;
  $\tuple{\States \uplus \States_\mathrm{new}, \Alphabet', \Trans', \InitialState} := \texttt{Eliminate}(M', \States)$\label{alg:StateElimination:Elim}\tcp*{MDP state elimination}
  $\Trans'' := \set{ \bot \mapsto \set{ \tuple{\tau, \Dirac{\bot}} } }$, $V(\bot) := 0$, $\mu' := \varnothing$\;
  \ForEach(\tcp*[f]{state merging}){$s_\mathrm{new} \in \States_\mathrm{new}$ and $\tuple{a, \mu} \in \Trans'(s)$\label{alg:StateElimination:MergeLoop}}{
    $\mu' := \mu' \cup \set{ \bot \mapsto \mu(s') \mid s' \in \support{\mu} \cap \States }$\;
    $\mu' := \mu' \cup \set{ s' \mapsto \mu(s'_\mathrm{new}) \mid s'_\mathrm{new} \in \support{\mu} \cap \States_\mathrm{new} }$\;
    $\Trans''(s_\mathrm{new}) := \Trans''(s_\mathrm{new}) \cup \set{ \tuple{a, \mu'} }$, $\mu' := \varnothing$\label{alg:StateElimination:StepIter}\;
  }
  $\texttt{StepBoundedVI}(V, \tuple{\Dom{\Trans''}, \Alphabet', \Trans'', \InitialState}, n, \mathit{opt})$\tcp*{step-bounded iteration}
}
\caption{\small Reward-bounded reachability via MDP state elimination}
\label{alg:StateElimination}
\end{algorithm}

\alg{elim} is shown as \Cref{alg:StateElimination}.
In line~\ref{alg:StateElimination:Elim}, the MDP state elimination procedure is called to eliminate all the regular states in \States.
As an extension to the schema of \Cref{fig:MDPStateElimination}, we also preserve the original outgoing transitions when we eliminate a relevant state (defined as in \Cref{sec:AlgorithmsSched}) because we need them in the next step:
In the loop starting in line~\ref{alg:StateElimination:MergeLoop}, we redirect
(1) all branches that go to non-new states to the added bottom state $\bot$ instead because they indicate that we can get stuck in an end component without reward, and
(2) all branches that go to new states to the corresponding original states instead.
This way, we merge the (absorbing, but not eliminated) new states with the corresponding regular (eliminated from incoming but not outgoing transitions) states.
Finally, in line~\ref{alg:StateElimination:StepIter}, the standard step-bounded value iteration is performed on the eliminated-merged MDP.
Again, the algorithm reduces the model \wrt simple schedulers such that one transition corresponds to a reward of~$1$; then the step-bounded algorithm makes the reward-positional choices.
As before, it also computes the probabilities for all bounds $i \leq n$ implicitly, which our implementation returns explicitly.

\begin{example}
\Cref{fig:ExampleMDPEliminated} shows our example MDP after state elimination, and \Cref{fig:ExampleMDPEliminatedMerged} shows the subsequent merged MDP.
For clarity, transitions to the same successor state distributions are shown in a combined way.
The evolution of the value vectors during step-bounded value iteration on the latter is shown in \Cref{tab:ExampleEliminatedVI}.
\end{example}

\begin{figure}[t]
\begin{minipage}[b]{.5\textwidth}
\centering
\begin{tikzpicture}[on grid,auto]
  \node[state] (s) {$s$};
  \coordinate[left=0.3 of s.west] (start);
  \node[state] (t) [right=1.5 of s] {$t$};
  \node[state] (v) [right=1.5 of t] {$v$};
  \node[state] (w) [below left=1.5 and 1.5 of s] {$w$};
  \node[lstate] (snew) [below=1.5 of s] {$\strut s_\mathrm{new}$};
  \node[lstate] (tnew) [right=1.5 of snew] {$\strut t_\mathrm{new}$};
  \node[lstate] (vnew) [right=1.5 of tnew] {$\strut v_\mathrm{new}$};
  \node[dot] (d1) [below=0.6 of s] {};
  \node[dot] (d2) [below right=0.6 and 0.33 of t] {};
  ;
  \path[-]
    (s) edge[bend left=10] node[swap,pos=0.85,inner sep=1pt] {\texttt{be}} (d1)
    (t) edge[bend left=10] node[swap,pos=0.1,inner sep=1pt] {\texttt{cbe}} (d1)
    (s) edge[bend right=10,line width=2pt,white] (d2)
    (s) edge[bend right=10] node[pos=0.05,inner sep=1pt] {\texttt{ad}} (d2)
    (t) edge[bend right=10] node[pos=0.75,inner sep=1pt] {\texttt{d}} (d2)
  ;
  \path[->]
    (start) edge node {} (s)
    (s) edge[bend left=30] node[pos=0.5,inner sep=1.5] {\texttt{aca}} (t)
    (v) edge[bend left=30] (vnew)
    (d1) edge[bend right=10] node[swap,pos=0.5,inner sep=0.5pt] {$0.5$} (w)
    (d1) edge[bend right=15] node[swap,pos=0.67,inner sep=1pt] {$0.25$} (snew)
    (d1) edge[bend right=20] node[pos=0.33,inner sep=0.5pt] {$0.25$} (tnew)
    (d2) edge[bend left=40] node[swap,pos=0.3,inner sep=1.5pt] {$0.8$} (tnew)
    (d2) edge[bend left=30] node[pos=0.15,inner sep=1pt] {$0.2$} (vnew)
    (t) edge[overlay,loop,out=-30,in=30,looseness=4] node[swap,pos=0.5,inner sep=1.5] {\texttt{ca}} (t)
    (w) edge[overlay,loop,out=210,in=150,looseness=4] (w)
    (snew) edge[loop,out=200,in=160,looseness=3,pos=0.5] (snew)
    (tnew) edge[loop,out=-20,in=20,looseness=3,pos=0.5] (tnew)
    (vnew) edge[loop,out=-20,in=20,looseness=3,pos=0.5,overlay] (vnew)
  ;
\end{tikzpicture}
\caption{$M^e_{\downarrow}$ after state elimination}
\label{fig:ExampleMDPEliminated}
\end{minipage}%
\begin{minipage}[b]{.5\textwidth}
\centering
\begin{tikzpicture}[on grid,auto]
  \node[state] (s) {$s$};
  \coordinate[left=0.3 of s.west] (start);
  \node[state] (t) [right=2.5 of s] {$t$};
  \node[state] (v) [below=1.5 of t] {$v$};
  \node[state] (w) [below=1.5 of s] {$\bot$};
  \node[dot] (d1) [below right=0.6 and 0.5 of s] {};
  \node[dot] (d2) [below left=0.6 and 0.5 of t] {};
  ;
  \path[->]
    (t) edge[out=200,in=20] node[pos=0.525,inner sep=1pt] {\texttt{ca}} (w)
  ;
  \path[-]
    (s) edge[bend left=20] node[swap,pos=0.5,inner sep=0.5pt] {\texttt{be}} (d1)
    (t) edge[bend right=10] node[swap,pos=0.15,inner sep=1pt] {\texttt{cbe}} (d1)
    (s) edge[bend left=10,line width=2pt,white] (d2)
    (s) edge[bend left=10] node[pos=0.15,inner sep=1pt] {\texttt{ad}} (d2)
    (t) edge[bend right=20] node[pos=0.5,inner sep=0.5pt] {\texttt{d}} (d2)
  ;
  \path[->]
    (start) edge node {} (s)
    (s) edge[bend right=35] node[swap,pos=0.5,inner sep=1pt] {\texttt{aca}} (w)
    (d1) edge[bend right=10] node[pos=0.6,inner sep=0.5pt] {$0.5$} (w)
    (d1) edge[out=-67,in=180,line width=2pt,white] (v)
    (d1) edge[out=-67,in=180] node[swap,pos=0.6,inner sep=1pt] {$0.25$} (v)
    (d1) edge[out=200,in=-110,looseness=2] node[pos=0.33,inner sep=1.5pt] {$0.5~$} (s)
    (d2) edge[bend right=10] node[swap,pos=0.6,inner sep=0.5pt] {$0.2$} (v)
    (d2) edge[out=-20,in=-70,looseness=2] node[swap,pos=0.33,inner sep=1.5pt] {$0.8~$} (t)
    (w) edge[overlay,loop,out=210,in=150,looseness=4] (w)
    (v) edge[overlay,loop,out=-30,in=30,looseness=4] (v)
  ;
\end{tikzpicture}
\caption{$M^e_{\downarrow}$ eliminated and merged}
\label{fig:ExampleMDPEliminatedMerged}
\end{minipage}
\end{figure}

\subsection{Correctness}
\label{sec:AlgorithmsCorrectness}

Let \Sched be a deterministic reward-positional scheduler for $\limit{M}{F}$.
It corresponds to a sequence of simple schedulers $\Sched_i$ for $\limit{M}{F}$ where $i \in \set{n, \ldots, 0}$ is the remaining reward that can be accumulated before the bound is reached.
For each state $s$ of $\limit{M}{F}$ and $i > 0$, each such $\Sched_i$ induces a (potentially substochastic) measure $\mu_s^i$ such that $\mu_s^i(s')$ is the probability to reach $s'$ from $s$ in $\limit{\limit{M}{F}}{\Sched_i}$ over paths whose last step has reward~$1$.
Let $\mu^0_s$ be the induced measure such that $\mu_s^0(s')$ is the probability under $\Sched_0$ to reach $s'$ without reward if it is a goal state and $0$ otherwise.
Using the recursion
$\overline{\mu}^i_s(s') \defeq \sum_{s''\in \States} \mu_s^{i} (s'') \cdot \overline{\mu}^{i-1}_{s''}(s')$ with $\overline{\mu}^0_s \defeq \mu^0_s$, the value $\overline{\mu}^n_s(s')$ is the probability to reach goal state $s'$ from $s$ in $\limit{M}{F}$ under~\Sched.
Thus we have $\max_{\Sched} \overline{\mu}^n_s(s') = \rewardreachprob{n}{\mathrm{max}}{F}$ and $\max_{\Sched} \overline{\mu}^n_s(s') = \rewardreachprob{n}{\mathrm{min}}{F}$ by \Cref{theo:SchedulerPower}.
If we distribute the maximum operation into the recursion, we get\vspace{-3mm}
\begin{equation}
\label{eq:MaxRecursive}
\max_{\Sched} \overline{\mu}^i_s(s') = \sum_{s''\in \States} \max_{\Sched_i} \mu_s^{i} (s'') \cdot \max_{\Sched} \overline{\mu}^{i-1}_{s''}(s')\vspace{-1mm}
\end{equation}
and an analogous formula for the minimum.
By computing extremal values \wrt simple schedulers for each reward step, we thus compute the value \wrt an optimal deterministic reward-positional scheduler for the bounded property overall.
The correctness of our algorithms now follows from the fact that they all implement precisely the right-hand side of \Cref{eq:MaxRecursive}, albeit in different ways:

$\overline{\mu}^0_s$ is always given as the initial value of $V$ as described at the very beginning of this section.
In \alg{modvi}, the call to \texttt{UnboundedVI} in loop iteration $i \geq 1$ then computes the optimal $\overline{\mu}^{i}_{\cdot}$ based on the relevant values for the optimal $\overline{\mu}^{i-1}_{\cdot}$ copied from the previous iteration.
In \alg{senum}, we enumerate the relevant measures $\mu_s^{\cdot}$ induced by all the simple schedulers as one transition each, then choose the optimal transition for each $i$ in the $i$-th iteration inside \texttt{StepBoundedVI}.
The argument for \alg{elim} is the same, the difference being that state elimination is what transforms all the measures into single transitions.

\section{Implementation}
\label{sec:Implementation}

\begin{figure}[t]
\centering
\includegraphics[width=0.89\textwidth]{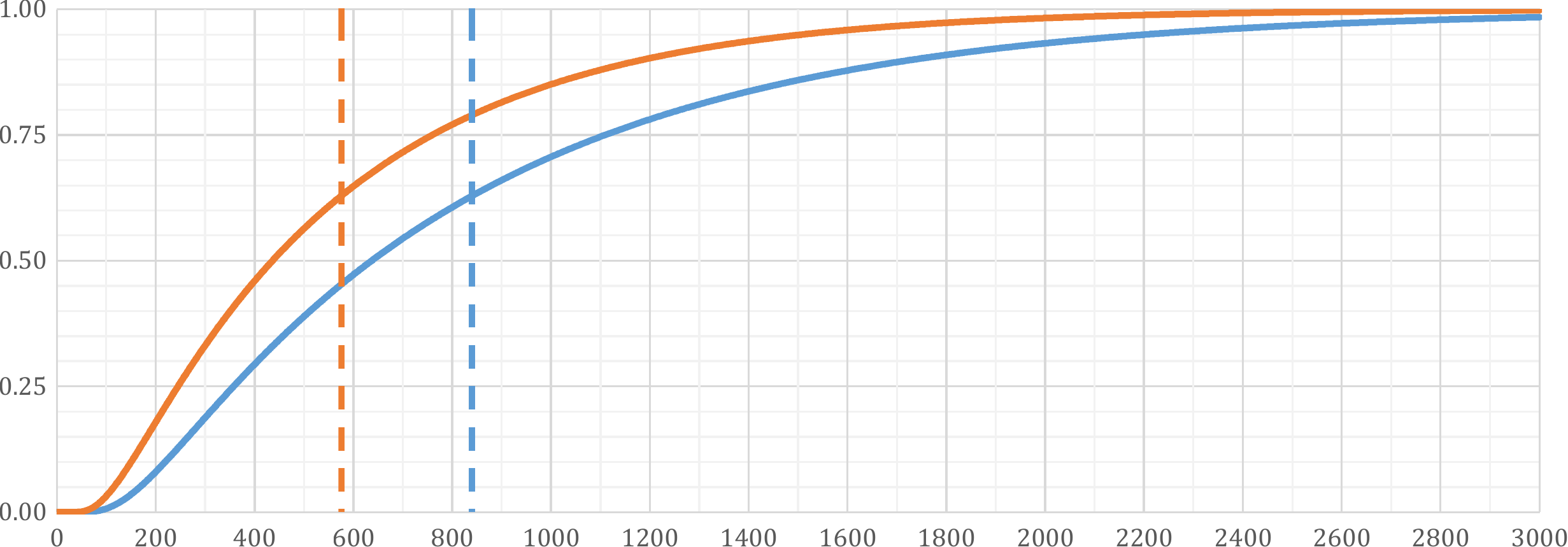}
\caption{Cdfs and means for the randomised consensus model ($H=6,K=4$)}
\label{fig:Graph}
\end{figure}

We have implemented the three new unfolding-free algorithms within \mcsta, the \toolset's stochastic timed systems model checker.
When asked to compute $\rewardreachprob{n}{\mathit{opt}}{\cdot}$, it also delivers \emph{all} values $\rewardreachprob{i}{\mathit{opt}}{\cdot}$ for $i \in \set{0, \dots, n}$ since the algorithms allow doing so at no overhead.
Instead of a single value, we thus get the entire (sub-)cdf.
Every single value is defined via an individual optimisation over schedulers.
However, we have seen in \Cref{sec:AlgorithmsCorrectness} that an optimal scheduler for bound $i$ can be extended to an optimal scheduler for $i+1$, so there exists an optimal scheduler for all bounds.
The max./min.\ cdf represents the probability distribution induced by that scheduler.
We show these functions for the randomised consensus case study~\cite{KNP12} in \Cref{fig:Graph}.
The top curve is the max.\ probability for the protocol to terminate within the number of coin tosses given on the x-axis.
The bottom curve is the min.\ probability.
For comparison, we also show the means $\reachreward{\mathrm{min}}{\cdot}$ and $\reachreward{\mathrm{max}}{\cdot}$ of these distributions as the left and right dashed lines, respectively.
Note that the min.\ expected value corresponds to the max.\ bounded probabilities and vice-versa.
As mentioned, using our new algorithms, it is now possible to compute the curves in the same amount of memory (and, in this case, also runtime) previously sufficient for the means~only.

We also implemented a convergence criterion to detect when the result will no longer increase for higher bounds.
Recall that \texttt{Un\-bounded\-VI} terminates when the max.\ error in the \emph{last} iteration is below $\epsilon$ (default $\epsilon = 10^{-6}$).
Our implementation considers the max.\ error $e$ over \emph{all} iterations of a call to \texttt{Un\-bounded\-VI} in \alg{modvi} or \texttt{Step\-Bounded\-VI} in \alg{senum} and \alg{elim}.
In the same spirit as the unbounded value iteration algorithm, we can then stop when $e < \epsilon$.
For the functions shown in \Cref{fig:Graph}, this happens at 4016 coin tosses for the max.\ and 5607 for the min.\ probability.

\section{Experiments}
\label{sec:Experiments}

\begin{table}[t]
\caption{State spaces}
\label{tab:StateSpaces}
\renewcommand{\arraystretch}{1.0}
\setlength{\tabcolsep}{2.6pt}
\centering
\begin{tabular}{ccc|rr|rrrr|rrr}
 & & & \multicolumn{2}{|c|}{unfolded} & \multicolumn{4}{|c|}{non-unfolded} & \multicolumn{3}{|c}{eliminated} \\
\multicolumn{2}{c}{model} & $b$ & states & time & states & trans & branch & time & states & trans & branch \\\hline
\multirow{7}{*}{\rotatebox[origin=c]{90}{BEB}}
& 4 & 58 & 114\sunit{k} & 2\sunit{s} & 2\sunit{k} & 3\sunit{k} & 4\sunit{k} & 0\sunit{s} & 1\sunit{k} & 1\sunit{k} & 3\sunit{k} \\
& 5 & 90 & 803\sunit{k} & 6\sunit{s} & 10\sunit{k} & 12\sunit{k} & 22\sunit{k} & 0\sunit{s} & 3\sunit{k} & 5\sunit{k} & 19\sunit{k} \\
& 6 & 143 & 5.9\sunit{M} & 35\sunit{s} & 45\sunit{k} & 60\sunit{k} & 118\sunit{k} & 0\sunit{s} & 16\sunit{k} & 28\sunit{k} & 104\sunit{k} \\
& 7 & 229 & 44.7\sunit{M} & 273\sunit{s} & 206\sunit{k} & 304\sunit{k} & 638\sunit{k} & 1\sunit{s} & 80\sunit{k} & 149\sunit{k} & 568\sunit{k} \\
& 8 & 371 & \multicolumn{2}{|c|}{\multirow{3}{*}{$>\!30\sunit{min}$}} & 1.0\sunit{M} & 1.6\sunit{M} & 3.4\sunit{M} & 3\sunit{s} & 0.4\sunit{M} & 0.8\sunit{M} & 3.1\sunit{M} \\
& 9 & 600 & & & 4.6\sunit{M} & 8.3\sunit{M} & 18.9\sunit{M} & 26\sunit{s} & 2.0\sunit{M} & 4.2\sunit{M} & 16.6\sunit{M} \\
& 10 & n/a & & & 22.2\sunit{M} & 44.0\sunit{M} & 102.8\sunit{M} & 138\sunit{s} & \multicolumn{3}{|c}{$>\!16\sunit{GB}$} \\\hline
\multirow{8}{*}{\rotatebox[origin=c]{90}{BRP}}
& 32,\,\phantom{0}6,\,2 & 179 & 9.4\sunit{M} & 40\sunit{s} & 0.1\sunit{M} & 0.1\sunit{M} & 0.1\sunit{M} & 0\sunit{s} & 0.1\sunit{M} & 0.4\sunit{M} & 7.1\sunit{M} \\
& 32,\,\phantom{0}6,\,4 & 347 & 50.2\sunit{M} & 206\sunit{s} & 0.2\sunit{M} & 0.2\sunit{M} & 0.2\sunit{M} & 1\sunit{s} & 0.2\sunit{M} & 1.0\sunit{M} & 20.3\sunit{M} \\
& 32,\,12,\,2 & 179 & 21.8\sunit{M} & 90\sunit{s} & 0.2\sunit{M} & 0.2\sunit{M} & 0.3\sunit{M} & 1\sunit{s} & 0.2\sunit{M} & 1.1\sunit{M} & 22.1\sunit{M} \\
& 32,\,12,\,4 & 347 & 122.0\sunit{M} & 499\sunit{s} & 0.6\sunit{M} & 0.7\sunit{M} & 0.7\sunit{M} & 2\sunit{s} & 0.6\sunit{M} & 3.2\sunit{M} & 62.0\sunit{M} \\
& 64,\,\phantom{0}6,\,2 & 322 & 38.2\sunit{M} & 157\sunit{s} & 0.1\sunit{M} & 0.2\sunit{M} & 0.2\sunit{M} & 0\sunit{s} & 0.1\sunit{M} & 1.3\sunit{M} & 53.8\sunit{M} \\
& 64,\,\phantom{0}6,\,4 & 630 & 206.9\sunit{M} & 826\sunit{s} & 0.4\sunit{M} & 0.4\sunit{M} & 0.5\sunit{M} & 1\sunit{s} & 0.4\sunit{M} & 3.8\sunit{M} & 153.7\sunit{M} \\
& 64,\,12,\,2 & 322 & 107.0\sunit{M} & 427\sunit{s} & 0.5\sunit{M} & 0.5\sunit{M} & 0.5\sunit{M} & 1\sunit{s} & 0.4\sunit{M} & 4.1\sunit{M} & 166.0\sunit{M} \\
& 64,\,12,\,4 & 630 & \multicolumn{2}{|c|}{$>\!30\sunit{min}$} & 1.3\sunit{M} & 1.4\sunit{M} & 1.5\sunit{M} & 4\sunit{s} & \multicolumn{3}{|c}{$>\!16\sunit{GB}$} \\\hline
\multirow{4}{*}{\rotatebox[origin=c]{90}{RCONS}}
& 4,\,4 & 2653 & 53.9\sunit{M} & 365\sunit{s} & 41\sunit{k} & 113\sunit{k} & 164\sunit{k} & 0\sunit{s} & 35\sunit{k} & 254\sunit{k} & 506\sunit{k} \\
& 4,\,8 & 7793 & \multicolumn{2}{|c|}{\multirow{3}{*}{$>\!30\sunit{min}$}} & 80\sunit{k} & 220\sunit{k} & 323\sunit{k} & 0\sunit{s} & 68\sunit{k} & 499\sunit{k} & 997\sunit{k} \\
& 6,\,2 & 2175 & & & 1.2\sunit{M} & 5.0\sunit{M} & 7.2\sunit{M} & 5\sunit{s} & 1.1\sunit{M} & 23.6\sunit{M} & 47.1\sunit{M} \\
& 6,\,4 & 5607 & & & 2.3\sunit{M} & 9.4\sunit{M} & 13.9\sunit{M} & 9\sunit{s} & 2.2\sunit{M} & 42.2\sunit{M} & 84.3\sunit{M} \\\hline
\multirow{4}{*}{\rotatebox[origin=c]{90}{CSMA}}
& 1 & 2941 & 31.2\sunit{M} & 276\sunit{s} & 13\sunit{k} & 13\sunit{k} & 13\sunit{k} & 0\sunit{s} & 13\sunit{k} & 13\sunit{k} & 15\sunit{k} \\
& 2 & 3695 & 191.1\sunit{M} & 1097\sunit{s} & 96\sunit{k} & 96\sunit{k} & 97\sunit{k} & 0\sunit{s} & 95\sunit{k} & 95\sunit{k} & 110\sunit{k} \\
& 3 & 5229 & \multicolumn{2}{|c|}{\multirow{2}{*}{$>\!30\sunit{min}$}} & 548\sunit{k} & 548\sunit{k} & 551\sunit{k} & 2\sunit{s} & 545\sunit{k} & 545\sunit{k} & 637\sunit{k} \\
& 4 & 8219 & & & 2.7\sunit{M} & 2.7\sunit{M} & 2.7\sunit{M} & 9\sunit{s} & 2.7\sunit{M} & 2.7\sunit{M} & 3.2\sunit{M} \\\hline
\multirow{2}{*}{\rotatebox[origin=c]{90}{FW}}
& short & 2487 & 8.8\sunit{M} & 150\sunit{s} & 4\sunit{k} & 6\sunit{k} & 6\sunit{k} & 0\sunit{s} & 4\sunit{k} & 111\sunit{k} & 413\sunit{k} \\
& long & 3081 & \multicolumn{2}{|c|}{$>\!30\sunit{min}$} & 0.2\sunit{M} & 0.5\sunit{M} & 0.5\sunit{M} & 1\sunit{s} & 0.2\sunit{M} & 2.4\sunit{M} & 7.7\sunit{M} \\
\end{tabular}
\end{table}

\begin{table}[t]
\caption{Performance}
\label{tab:Performance}
\renewcommand{\arraystretch}{1.0}
\setlength{\tabcolsep}{2.5pt}
\centering
\begin{tabular}{ccc|rrrr|rr|rrrr}
 & & & \multicolumn{4}{|c|}{\alg{modvi}} & \multicolumn{2}{|c|}{\alg{senum}} & \multicolumn{4}{|c}{\alg{elim}} \\
\multicolumn{2}{c}{model} & $b$ & iter & \# & mem & rate & enum & mem & elim & iter & mem & rate \\\hline
\multirow{7}{*}{\rotatebox[origin=c]{90}{BEB}}
& 4 & 58 & 0\sunit{s} & 257 & 40\sunit{M} & $\infty\funit{1}{s}$ & 0\sunit{s} & 41\sunit{M} & 0\sunit{s} & 0\sunit{s} & 40\sunit{M} & $\infty\funit{1}{s}$ \\
& 5 & 90 & 0\sunit{s} & 407 & 40\sunit{M} & 850\funit{1}{s} & 0\sunit{s} & 45\sunit{M} & 0\sunit{s} & 0\sunit{s} & 48\sunit{M} & $\infty\funit{1}{s}$ \\
& 6 & 143 & 1\sunit{s} & 651 & 54\sunit{M} & 223\funit{1}{s} & 1\sunit{s} & 65\sunit{M} & 0\sunit{s} & 0\sunit{s} & 107\sunit{M} & 1430\funit{1}{s} \\
& 7 & 229 & 6\sunit{s} & 1055 & 127\sunit{M} & 38\funit{1}{s} & 11\sunit{s} & 210\sunit{M} & 2\sunit{s} & 0\sunit{s} & 409\sunit{M} & 1145\funit{1}{s} \\
& 8 & 371 & 49\sunit{s} & 1714 & 345\sunit{M} & 7\funit{1}{s} & 88\sunit{s} & 588\sunit{M} & 12\sunit{s} & 2\sunit{s} & 1.5\sunit{G} & 247\funit{1}{s} \\
& 9 & 600 & 425\sunit{s} & 2769 & 1.7\sunit{G} & 1\funit{1}{s} & 960\sunit{s} & 2.6\sunit{G} & 67\sunit{s} & 14\sunit{s} & 6.6\sunit{G} & 43\funit{1}{s} \\
& 10 & n/a & \multicolumn{4}{|c|}{$>\!30\sunit{min}$} & \multicolumn{2}{|c|}{$>\!30\sunit{min}$} & \multicolumn{4}{|c}{$>\!16\sunit{GB}$} \\\hline
\multirow{8}{*}{\rotatebox[origin=c]{90}{BRP}}
& 32,\,\phantom{0}6,\,2 & 179 & 1\sunit{s} & 803 & 55\sunit{M} & 256\funit{1}{s} & 160\sunit{s} & 474\sunit{M} & 4\sunit{s} & 1\sunit{s} & 775\sunit{M} & 164\funit{1}{s} \\
& 32,\,\phantom{0}6,\,4 & 347 & 4\sunit{s} & 1419 & 85\sunit{M} & 102\funit{1}{s} & 498\sunit{s} & 1.2\sunit{G} & 12\sunit{s} & 7\sunit{s} & 2.6\sunit{G} & 61\funit{1}{s} \\
& 32,\,12,\,2 & 179 & 3\sunit{s} & 803 & 90\sunit{M} & 85\funit{1}{s} & 569\sunit{s} & 1.3\sunit{G} & 13\sunit{s} & 4\sunit{s} & 2.8\sunit{G} & 56\funit{1}{s} \\
& 32,\,12,\,4 & 347 & 13\sunit{s} & 1419 & 196\sunit{M} & 33\funit{1}{s} & 1467\sunit{s} & 3.5\sunit{G} & 40\sunit{s} & 22\sunit{s} & 6.1\sunit{G} & 20\funit{1}{s} \\
& 64,\,\phantom{0}6,\,2 & 322 & 3\sunit{s} & 1414 & 76\sunit{M} & 129\funit{1}{s} & \multicolumn{2}{|c|}{\multirow{4}{*}{$>\!30\sunit{min}$}} & 31\sunit{s} & 16\sunit{s} & 4.7\sunit{G} & 24\funit{1}{s} \\
& 64,\,\phantom{0}6,\,4 & 630 & 17\sunit{s} & 2605 & 137\sunit{M} & 45\funit{1}{s} & & & 114\sunit{s} & 91\sunit{s} & 14.1\sunit{G} & 8\funit{1}{s} \\
& 64,\,12,\,2 & 322 & 10\sunit{s} & 1414 & 146\sunit{M} & 40\funit{1}{s} & & & 132\sunit{s} & 51\sunit{s} & 13.9\sunit{G} & 8\funit{1}{s} \\
& 64,\,12,\,4 & 630 & 50\sunit{s} & 2605 & 318\sunit{M} & 15\funit{1}{s} & & & \multicolumn{4}{|c}{$>\!16\sunit{GB}$} \\\hline
\multirow{4}{*}{\rotatebox[origin=c]{90}{RCONS}}
& 4,\,4 & 2653 & 37\sunit{s} & 21728 & 62\sunit{M} & 124\funit{1}{s} & 2\sunit{s} & 126\sunit{M} & 1\sunit{s} & 3\sunit{s} & 224\sunit{M} & 1842\funit{1}{s} \\
& 4,\,8 & 7793 & 222\sunit{s} & 66704 & 85\sunit{M} & 65\funit{1}{s} & 4\sunit{s} & 187\sunit{M} & 2\sunit{s} & 16\sunit{s} & 384\sunit{M} & 933\funit{1}{s} \\
& 6,\,2 & 2175 & 1383\sunit{s} & 19136 & 679\sunit{M} & 2\funit{1}{s} & \multicolumn{2}{|c|}{\multirow{2}{*}{$>\!30\sunit{min}$}} & 136\sunit{s} & 169\sunit{s} & 11.9\sunit{G} & 20\funit{1}{s} \\
& 6,\,4 & 5607 & \multicolumn{4}{|c|}{$>\!30\sunit{min}$} & & & 275\sunit{s} & 879\sunit{s} & 13.4\sunit{G} & 11\funit{1}{s} \\\hline
\multirow{4}{*}{\rotatebox[origin=c]{90}{CSMA}}
& 1 & 2941 & 4\sunit{s} & 11904 & 45\sunit{M} & 1548\funit{1}{s} & 5\sunit{s} & 46\sunit{M} & 0\sunit{s} & 0\sunit{s} & 60\sunit{M} & $\infty\funit{1}{s}$ \\
& 2 & 3695 & 23\sunit{s} & 15061 & 66\sunit{M} & 312\funit{1}{s} & \multicolumn{2}{|c|}{\multirow{3}{*}{$>\!30\sunit{min}$}} & 1\sunit{s} & 3\sunit{s} & 190\sunit{M} & 2437\funit{1}{s} \\
& 3 & 5229 & 170\sunit{s} & 21426 & 185\sunit{M} & 62\funit{1}{s} & & & 3\sunit{s} & 24\sunit{s} & 839\sunit{M} & 429\funit{1}{s} \\
& 4 & 8219 & 1270\sunit{s} & 33538 & 606\sunit{M} & 13\funit{1}{s} & & & 19\sunit{s} & 192\sunit{s} & 3.8\sunit{G} & 86\funit{1}{s} \\\hline
\multirow{2}{*}{\rotatebox[origin=c]{90}{FW}}
& short & 2487 & 1\sunit{s} & 6012 & 40\sunit{M} & 2072\funit{1}{s} & 29\sunit{s} & 79\sunit{M} & 0\sunit{s} & 1\sunit{s} & 113\sunit{M} & 3109\funit{1}{s} \\
& long & 3081 & 51\sunit{s} & 9431 & 107\sunit{M} & 60\funit{1}{s} & 205\sunit{s} & 880\sunit{M} & 13\sunit{s} & 23\sunit{s} & 1.4\sunit{G} & 132\funit{1}{s} \\
\end{tabular}
\end{table}

We use five case studies from the literature to evaluate the applicability and performance of our three algorithms and their implementation:
\begin{itemize}
\item \textbf{BEB}~\cite{GDF09}:
MDP of a bounded exponential backoff procedure with max.\ backoff value $K=4$ and $H \in \set{4, \dots, 10}$ parallel hosts.
We compute the max.\ probability of any host seizing the line while all hosts enter backoff $\leq b$~times.
\item \textbf{BRP}~\cite{HH09}:
The PTA model of the bounded retransmission protocol with $N \in \set{ 32, 64 }$ frames to transmit, retransmission bound $\mathit{MAX} \in \set{6, 12}$ and transmission delay $\mathit{TD} \in \set{ 2, 4 }$ time units.
We compute the max.\ and min.\ probability that the sender reports success in $\leq b$ time units.
\item \textbf{RCONS}~\cite{KNP12}:
The randomised consensus shared coin protocol MDP as described in \Cref{sec:Implementation} for $N \in \set{ 4, 6 }$ parallel processes and constant $K \in \set{ 2, 4, 8 }$.
\item \textbf{CSMA}~\cite{HH09}:
PTA model of a communication protocol using CSMA/CD, with max.\ backoff counter $\mathit{BCMAX} \in \set{ 1, \dots, 4 }$.
We compute the min.\ and max.\ probability that both stations deliver their packets by deadline $b$ time units.
\item \textbf{FW}~\cite{KNP12}:
PTA model (``Impl'' variant) of the IEEE 1394 FireWire root contention protocol with either a short or a long cable.
We ask for the min.\ probability that a leader (root) is selected before time bound $b$.
\end{itemize}
Experiments were performed on an Intel Core i5-6600T system (2.7\sunit{GHz}, 4~cores) with 16\sunit{GB} of memory running 64-bit Windows~10 and a
timeout of 30~minutes.

If we look back at the description of the algorithms in \Cref{sec:Algorithms}, we see that the only extra states introduced by \alg{modvi} compared to checking an unbounded probabilistic reachability or expected-reward property are the new states $s_\mathrm{new}$.
However, this can be avoided in the implementation by checking for reward-one branches on-the-fly.
The transformations performed in \alg{senum} and \alg{elim}, on the other hand, will reduce the number of states, but may add transitions and branches.
\alg{elim} may also create large intermediate models.
In contrast to \alg{modvi}, these two algorithms may thus run out of memory even if unbounded properties can be checked.
In \Cref{tab:StateSpaces}, we show the state-space sizes
(1) for the traditional unfolding approach (``unfolded'') for the bound $b$ where the values have converged,
(2) when unbounded properties are checked or \alg{modvi} is used (``non-unfolded''), and
(3) after state elimination and merging in \alg{elim}.
We report thousands (\ssunit{k}) or millions (\ssunit{M}) of states, transitions (``trans'') and branches (``branch'').
The values for \alg{senum} are the same as for \alg{elim}.
Times are for the state-space exploration phase only, so the time for ``non-unfolded'' will be incurred by all three new algorithms.
We see that avoiding unfolding is a drastic reduction.
In fact, 16\sunit{GB} of memory are not sufficient for the larger unfolded models, so we had to enable \mcsta's disk-based technique~\cite{HH15}.
State elimination leads to an increase in transitions and especially branches, drastically so for BRP, the exception being BEB.

In \Cref{tab:Performance}, we report the performance results for all three new algorithms when run until the values have converged at bound value~$b$.
For \alg{senum}, we used the variant based on value iteration since it consistently performed better than the one using DTMC state elimination.
``iter'' denotes the time needed for (unbounded or step-bounded) value iteration, while ``enum'' and ``elim'' are the times needed for scheduler enumeration resp.\ state elimination and merging.
``\#''~is the total number of outer-loop iterations performed during the calls to \texttt{UnboundedVI}.
``rate'' is the number of bound values computed per second.
Memory usage in columns ``mem'' is \mcsta's peak working set, including state space exploration, reported in mega- (\ssunit{M}) or gigabytes (\ssunit{G}).
\mcsta is garbage-collected, so these values are higher than necessary since full collections only occur when the system runs low on memory.
The values related to value iteration for \alg{senum} are the same as for \alg{elim}.
In general, we see that \alg{senum} uses less memory than \alg{elim}, but is much slower in all cases.
If it works and does not blow up the model too much, \alg{elim} is significantly faster than \alg{modvi}, making up for the time spent on state elimination with much faster value iteration rates.

\section{Conclusion}
\label{sec:Conclusion}

We presented three algorithms to model-check cost-/reward-bounded properties on MDP without unfolding.
In contrast to recent related work similar to our first algorithm~\cite{BDFK16,HHS16}, we also consider the application to time-bounded properties on PTA.
By avoiding unfolding and returning the entire probability distribution at no extra cost, our techniques could finally make cost-bounded probabilistic (timed) model checking feasible in practical applications.

\paragraph{Outlook.}
The digital clocks approach for PTA was considered the most limited in scalability.
Our techniques lift some of its most significant practical limitations.
Moreover, time-bounded analysis without unfolding and with computation of the entire distribution in this manner is not feasible for the traditionally more scalable zone-based approaches~\cite{NPS13} because zones abstract from concrete timing.
We see the possibility to improve the state elimination approach by removing transitions that are linear combinations of others and thus unnecessary.
This may reduce the transition and branch blowup on models like the BRP case.

Our algorithms \alg{senum} and \alg{elim} can be extended to \emph{long-run average} properties.
For a random variable $X_i$ for the reward obtained in each step, their value is defined as
(A)~$\expm{\Sched}{M}(\lim_{n\to\infty} ({\sum_{i=0}^{n} X_{i}'})/({n+1}))$.
This allows to express \eg the average energy usage or the expected amount of time spent in some system state.
Note this is only appropriate if one transition corresponds to one abstract time unit, which is not the case in a digital clocks MDP.
In practice, also \emph{long-run reward-average} properties are of interest, \eg average energy consumption per subtask performed.
They can be expressed by using a second reward variable $Y_i$ and considering
(B)~$\expm{\Sched}{M}( \lim_{n\to\infty} ({\sum_{i=0}^{n} X_{i}})/({\sum_{i=0}^{n} Y_{i}}))$.
Previous work on solving (B) uses graph decomposition and linear programming~\cite{Alf98,BA95,EJ11}.
Using the ideas of  \alg{senum} and \alg{elim}, we can transform this problem into (A), for which the policy iteration-based algorithm of Howard and Veinott~\cite{How60,Put94,Vei66} is applicable.

\bibliography{paper}{}
\bibliographystyle{splncs03}

\end{document}